\documentstyle[12pt]{article}

\catcode`\@=11
\long\def\@makefntext#1{
\protect\noindent \hbox to 3.2pt {\hskip-.9pt
$^{{\ninerm\@thefnmark}}$\hfil}#1\hfill}                

\def\@makefnmark{\hbox to 0pt{$^{\@thefnmark}$\hss}}  
\def\ps@myheadings{\let\@mkboth\@gobbletwo
\def\@oddhead{\hbox{}
\rightmark\hfil\ninerm\thepage}
\def\@oddfoot{}\def\@evenhead{\ninerm\thepage\hfil
\leftmark\hbox{}}\def\@evenfoot{}
\def\sectionmark##1{}\def\subsectionmark##1{}}

\renewcommand{\thefootnote}{\fnsymbol{footnote}}

\newcounter{sectionc}\newcounter{subsectionc}\newcounter{subsubsectionc}
\renewcommand{\section}[1] {\vspace*{0.6cm}\addtocounter{sectionc}{1}
\setcounter{subsectionc}{0}\setcounter{subsubsectionc}{0}\noindent
        {\normalsize\bf\thesectionc. #1}\par\vspace*{0.4cm}}
\renewcommand{\subsection}[1] {\vspace*{0.6cm}\addtocounter{subsectionc}{1}
        \setcounter{subsubsectionc}{0}\noindent
        {\normalsize\it\thesectionc.\thesubsectionc. #1}\par\vspace*{0.4cm}}
\renewcommand{\subsubsection}[1]
{\vspace*{0.6cm}\addtocounter{subsubsectionc}{1}
        \noindent
{\normalsize\rm\thesectionc.\thesubsectionc.\thesubsubsectionc.
        #1}\par\vspace*{0.4cm}}

\newcounter{appendixc}
\newcounter{subappendixc}[appendixc]
\newcounter{subsubappendixc}[subappendixc]

\renewcommand{\appendix}[1] {\vspace*{0.6cm}
        \refstepcounter{appendixc}
        \setcounter{figure}{0}
        \setcounter{table}{0}
        \setcounter{equation}{0}
        \renewcommand{\thefigure}{\Alph{appendixc}.\arabic{figure}}
        \renewcommand{\thetable}{\Alph{appendixc}.\arabic{table}}
        \renewcommand{\theappendixc}{\Alph{appendixc}}
        \renewcommand{\theequation}{\Alph{appendixc}.\arabic{equation}}
        \noindent{\bf Appendix \theappendixc #1}\par\vspace*{0.4cm}}

\def\abstracts#1{{
\centering{\begin{minipage}{12.2truecm}\vspace*{.1cm}
        \footnotesize\baselineskip=12pt\noindent
        \parindent=0pt #1
        \end{minipage}}\par}}

\renewenvironment{thebibliography}[1]
        {\begin{list}{\arabic{enumi}.}
        {\usecounter{enumi}\setlength{\parsep}{0pt}
\setlength{\leftmargin 1.25cm}{\rightmargin 0pt}
         \setlength{\itemsep}{0pt} \settowidth
        {\labelwidth}{#1.}\sloppy}}{\end{list}}

\newcounter{itemlistc}
\newcounter{romanlistc}
\newcounter{alphlistc}
\newcounter{arabiclistc}

\newcommand{\fcaption}[1]{
        \refstepcounter{figure}
        \setbox\@tempboxa = \hbox{\footnotesize Fig.~\thefigure. #1}
        \ifdim \wd\@tempboxa > 6in
           {\begin{center}
        \parbox{6in}{\footnotesize\baselineskip=12pt Fig.~\thefigure. #1}
            \end{center}}
        \else
             {\begin{center}
             {\footnotesize Fig.~\thefigure. #1}
              \end{center}}
        \fi}

\newcommand{\tcaption}[1]{
        \refstepcounter{table}
        \setbox\@tempboxa = \hbox{\footnotesize Table~\thetable. #1}
        \ifdim \wd\@tempboxa > 6in
           {\begin{center}
        \parbox{6in}{\footnotesize\baselineskip=12pt Table~\thetable. #1}
            \end{center}}
        \else
             {\begin{center}
             {\footnotesize Table~\thetable. #1}
              \end{center}}
        \fi}

\def\@citex[#1]#2{\if@filesw\immediate\write\@auxout
        {\string\citation{#2}}\fi
\def\@citea{}\@cite{\@for\@citeb:=#2\do
        {\@citea\def\@citea{,}\@ifundefined
        {b@\@citeb}{{\bf ?}\@warning
        {Citation `\@citeb' on page \thepage \space undefined}}
        {\csname b@\@citeb\endcsname}}}{#1}}

\newif\if@cghi
\def\cite{\@cghitrue\@ifnextchar [{\@tempswatrue
        \@citex}{\@tempswafalse\@citex[]}}
\def\citelow{\@cghifalse\@ifnextchar [{\@tempswatrue
        \@citex}{\@tempswafalse\@citex[]}}
\def\@cite#1#2{{$\null^{#1}$\if@tempswa\typeout
        {IJCGA warning: optional citation argument
        ignored: `#2'} \fi}}

 1
 1
 1

\font\ninerm=cmr9

\begin{document}
\begin{flushright}
NSF-ITP-95-110\\
NUB-TH-3127/95\\
CTP-TAMU-34/95\\
\end{flushright}
\centerline{\normalsize\bf SUPERSYMMETRIC DARK MATTER$\footnote{Talk at the
SUSY95 Workshop at Ecole Polytechnique,Palaiseau,May 1995. It is a shortened
version of a similar talk given at the Pascos/Hopkins
Workshop at the John Hopkins, Baltimore,Maryland, March 1995.}$}
\baselineskip=22pt

\centerline{\footnotesize PRAN NATH}

\baselineskip=13pt
\centerline{$\footnote { Permanent Address}$\footnotesize\it ~Department of
Physics, Northeastern University,}
\baselineskip=12pt
\centerline{\footnotesize\it Boston, MA  02119/USA}
\centerline{\footnotesize and }
\baselineskip=13pt
\centerline{\footnotesize\it Institute for Theoretical Physics}
\baselineskip=12pt
\centerline{\footnotesize\it University of California,Santa Barbara,CA 93117
/USA}
\centerline{\footnotesize E-mail: nath@itp.ucsb.edu}
\vspace*{0.5cm}
\centerline{\footnotesize R. ARNOWITT}
\baselineskip=13pt
\centerline{\footnotesize\it Center for Theoretical Physics,Department of
Physics,}
\baselineskip=12pt
\centerline{\footnotesize\it Texas A\& M University,College
Station,TX77843/USA}
\centerline{\footnotesize E-mail: arnowitt@phys.tamu.edu}

\vspace*{0.9cm}
\abstracts{A review of
supersymmetric dark matter in minimal supergravity unification with
R-parity invariance and with radiative breaking  of the electro-weak symmetry
is given. The analysis  shows the lightest neutralino is the LSP over most
of the parameter space of the supergravity model. The event rates in
neutralino-nucleus scattering in dark matter detectors are also discussed.
It is found that the event rates are sensititive to the constraint  from
the $b\rightarrow s\gamma$ experiment.It is also found that the event rates
are sensitive to the constraints of relic density and in our anaysis we have
used the accurate method for
the computation of the neutralino relic density.
  Finally,the effect of the new results on quark polarizabilities,  from the
 data of the Spin Muon Collaboration, on event rates is also discussed.
 The analysis  shows that the event
rates for the Ge detectors and for other detectors which use  heavy targets are
only
negligibly affected.}

\normalsize\baselineskip=15pt
\setcounter{footnote}{0}
\renewcommand{\thefootnote}{\alph{footnote}}

\section{Introduction}
Considerable evidence for the presence of dark
matter in the universe exists: in our galaxy, in other galaxies and in galactic
clusters. The rotation curves of luminous matter
in spiral galaxies, point to massive amounts of non-luminous matter in
the halo of galaxies and provide perhaps the strongest evidence for the
existence of dark matter$^1$.  There are many possible candidates for
 dark matter both in particle physics and in astronomy. Thus in astronomy
possible candidates for dark matter are  Jupiters,
brown dwarfs, neutron stars, black holes etc, while in particle physics
one has the possibility of axions, neutrinos,sneutrinos, neutralinos etc,
An important constraint arises from the fact that not
all of the dark matter in the Universe can be baryonic in nature.  First,
the baryonic dark matter is constrained severely from analysis of
nucleosynthesis
which show $\Omega_B\leq 0.1$. Second, the recent results of the MACHO
Collaboration$^2$, and from EROS$^3$
indicate that at best only $20-30\%$ of the halo of galaxies is composed of
MACHO'S
(Massive Compact Halo Objects), and thus the remainder must be non-baryonic
dark matter (NBDM).
The non-baryonic dark matter could be either hot (HDM) or cold (CDM).
The HDM could be one of the neutrino species (most likely possibility is the
tau
neutrino $\nu_{\tau}$), while the CDM could be either an axion, or a SUSY
particle$^4$
(a sneutrino $\tilde\nu$ or a neutralino  $\chi$).  In the following we shall
pursue
the SUSY possibility and assume that CDM is either a $\tilde\nu$ or a $\chi$.
Actually it
turns out that in supergravity unification$^5$, with radiative breaking of the
electro-weak symmetry, the lightest neutralino turns out to be the LSP over
most of the parameter space of the model$^{6}$.  Thus the model actually
predicts the
neutralino to be the CDM$^{6}$.  Further, we assume a mix of cold and hot dark
matter
in the ratio of 2:1 as indicated by the COBE data.  The quantity that appears
in theoretical analyses is $\Omega h^2$, where h is the Hubble parameter in
units
 of $ 100Km/sMpc$.
Currently the experimental uncertainty in h is given by$^7$
$0.82 \pm 0.17$ (Freedman~ et. al) ; $0.53 \pm 0.05$ (Sandage~ et. al).
For the purpose of our analyses here we shall assume an h in the range
$0.4\leq h \leq 0.8$ consistent with the above data.  Then assuming
that $\Omega_B=0.1$, and $\Omega_{CDM}:\Omega_{HDM}$=2:1 one finds that

\begin{equation}
0.1\leq\Omega h^2 \leq 0.4
\end{equation}

As mentioned above there are two neutral states, the lightest neutralino
 $\chi$ and
the sneutrino $\tilde\nu$, which are possible candidates for CDM in
supersymmetric
models, for example, the MSSM.  However, in the MSSM there are many arbitrary
parameters and the model does not predict what the LSP is. The situation is
radically different in supergravity unified models$^5$.  Here the parameter
space is
five dimensional and reduces to a four dimensional space after fixing the
Z-mass
using radiative breaking of the electro-weak symmetry.  The parameter space of
the model is then characterized by
%
$m_0,m_{1/2},A_0,tan\beta$
where $m_0$ is the universal scalar mass, $m_{1/2}$ is the universal gaugino
mass,
 $A_0$ is the universal trilinear coupling at the GUT Scale and $tan\beta=
v_2/v_1$ where
$v_2$ gives mass to the top quark and $v_1$ gives mass to the bottom quark.  We
limit
the parameter space by the fine tuning criterion
%
$m_0,m_{gluino}\leq 1 TeV$
\noindent
where  $m_{gluino}=(\alpha_3/\alpha_G) m_{1/2}$ and also limit
$tan\beta$ (since $tan\beta$ is the ratio of two VeVs, a
large $tan\beta$ also implies a finetuning) so that $tan\beta\leq 20$.
 In this domain, we find
that the lightest neutralino is also the LSP over most of the allowed region.

\section{Neutralino Relic Density}
%

The neutralino relic density at current temperatures is given by$^{8}$
\begin{equation}
\Omega h{_0}{^2}=4.75\times 10^{-40}(T_{\tilde Z_1}/T_{\gamma})^3
(T_{\gamma}/2.75)^3 N_F^{1/2}
(GeV^{-2}/J(x_f))g/cm^3
\end{equation}
\noindent
Here $T_{\gamma}$ is the current photon temperature,$n_F$ is the
 effective degrees of
freedom computed at the freeze-out temperature,
 $(T_{\chi}/T_{\gamma})^3$
is a reheating factor,and $J(x_f)$ is given by
\begin{equation}
J(x_f)=\int_{0}^{x_f} <\sigma v> dx
\end{equation}
\noindent
\noindent
$<\sigma v>$ is the thermal average of the neutralino annihilation cross
section and $x\equiv kT/m_{\tilde{Z}_1}$.
Now J receives contributions from Z-exchange, Higgs exchange and from the
s-fermion exchange in the t-channel.
%
\noindent
For the s-fermion exchange the conventional approximation$^{8}$ of expanding
 $<\sigma v>=
a+bv^2$
in the integrand in eq. (3) is valid and we use this approximation.
However, for the Z and Higgs pole terms such an expansion is a poor
approximation in the region below the poles$^{9-12}$.  Thus for $J_{Higgs}$ and
$J_{Z-pole}$ we use
a rigorous thermal averaging over the poles.  For example, consider the
annihilation via the lightest Higgs pole.  Here
\begin{eqnarray}
 \sigma v=\frac {A_{Higgs}}{m_{\tilde{Z}_1^4}}\frac {v^2}{((v^2-\epsilon_h)^2
+\gamma_h^2))}\\
\epsilon_h=(m_h^2-4m_{\tilde{Z}_1^2}/m_{\tilde{Z}_1^2}\\
\gamma_h=m_h \Gamma_h/m_{\tilde{Z}_1^2}
\end{eqnarray}
where $m_h$ is the Higgs mass, and  $\Gamma_h$  is the Higgs width.


Computationally eq. (3) implies a double integration over a
pole.  Since the pole is associated with a  very small width numerical
integrations are tricky as a sharp resonance can be easily missed.  A more
reliable procedure is to reduce eq. (3) to a single integral so that$^{10-11}$
\begin{equation}
J_{Higgs}(x_f)=\frac {A_{Higgs}}{2\sqrt 2 m_{\tilde Z_1^4}}
[I_{1h}+\frac {\epsilon_h}{\gamma_h} I_{2h}]
\end{equation}
\begin{eqnarray}
I_{1h}=\frac {1}{2}\int_{0}^{\infty}dyy^{-\frac{1}{2}}e^{-y}
\log [\frac {(4yx_f-\epsilon_h)^2+\gamma_h^2}{\epsilon_h^2+\gamma_h^2}]\\
I_{2h}=\frac {1}{2}\int_{0}^{\infty}dyy^{-\frac{1}{2}}e^{-y}
[tan^{-1} (\frac {(4yx_f-\epsilon_h)^2+\gamma_h^2}{\gamma_h})
+tan^{-1}(\frac {\epsilon_h}{\gamma_h})]
\end{eqnarray}
\noindent
A similar analysis can be carried out for $J_Z$.  What one finds then is
that  eq. (7) and the similar expression for $J_Z$ give
a smooth result on integration over the poles.

\section{Analytic Analysis of the Neutralino Composition}

 The lightest neutralino $\tilde Z_1$ is a linear combination of the four
neutral states
$(\tilde W,\tilde B,\tilde H_1,\tilde H_2)$ so that
\begin{equation}
\tilde Z_1 = n_1\tilde W_3 +  n_2\tilde B + n_3 \tilde H_1^0 + n_4
\tilde H_2^0~
\end{equation}
\noindent
where $n_i(i=1-4)$ are the co-efficient of the components
of the $\tilde Z_1$ eigenvector and are discussed below.
%
The co-efficients $n_i$ play an
important role both in the relic density analyses as well as in the analyses of
event rates in neutralino-nucleus scattering.  It is useful then to gain an
analytic understanding of the parametric dependence of $n_i$ on the basic
paramters
of the model.  The $n_i$ are determined by the neutralino mass matrix

\begin{equation}
M_{\tilde{Z}}=\pmatrix{\offinterlineskip
{\tilde{m}_{2}}&o&\vrule\strut&a&b\cr
o&{\tilde{m}_{1}}&\vrule\strut&c&d\cr
\noalign{\hrule}
a&c&\vrule\strut&o&{-\mu}\cr
b&d&\vrule\strut&{-\mu}&o\cr}
\end{equation}

\noindent
where $\tilde{m}_{i}=(\alpha_{i}/\alpha_{3})m_{\tilde{g}},~ a=M_{Z}
cos\theta_{W}cos\beta,~ b=-M_{Z}cos\theta_{W}sin\beta,~
c=-M_{Z}sin\theta_{W}cos\beta$ and $d=M_{Z}sin\theta_{W}sin\beta$.
Here $\tilde m_2(\tilde m_1)$ are the SU(2)(U(1)) gaugino masses determined by
 the relation $\tilde m_i=(\alpha_i/\alpha_G)m_{1/2}$.
   The parameter $\mu$ is determined by fixing the Z-mass using
radiative breaking of the electro-weak symmetry.  It is found that over most of
the parameter space $\mu$  is determined to be large i.e. $|\mu^2/M_Z^2|>>1$.
In this
domain one can carry out a perturbative expansion of $n_i$ in
$M_Z/\mu$. One finds that
to second order in ($M_Z/\mu)$ one has$^{13}$
\begin{eqnarray}
n_{1} & \cong &-{1\over2}{M_{Z}\over\mu}{1\over  {(1-\tilde{m}_{1}^{2}/\mu^{2}
)}}
{M_{Z}\over{\tilde{m}_{2}}-\tilde{m}_{1}}
sin2\theta_{W}\left[sin2\beta +{\tilde{m}_{1}\over\mu}\right]\\
n_{2} & = &1-{1\over2}{M_{Z}^{2}\over\mu^{2}}{1\over
 {(1-\tilde{m}_{1}^{2}/\mu^{2} )}}
sin^{2}\theta_{W}\left[1+{\tilde{m}_{1}\over\mu} sin2\beta
+{{\tilde{m}_{1}^{2}}\over
\mu^{2}}\right]\\
n_{3} & = &{M_{Z}\over\mu}{1\over
 {(1-\tilde{m}_{1}^{2}/\mu^{2} )}}
sin\theta_{W}sin\beta\left[1+{\tilde{m}_{1}\over\mu} ctn\beta\right]\\
n_{4} & = &-{M_{Z}\over\mu}{1\over
 {(1-\tilde{m}_{1}^{2}/\mu^{2} )}}
sin\theta_{W} cos\beta\left[1+{\tilde{m}_{1}\over\mu} tan\beta\right]
\end{eqnarray}
\noindent
The expansion of eqs.(12-15) is found to be accurate to $(3-5)\%$
 over a significant
region of the parameter space.  From the above one can easily see that
%
$n_2>n_1,n_3,n_4, |n_3|>|n_4|$
where in getting this result we have used the radiative electro-weak symmetry
breaking relation $tan\beta>1$.  We note that eq. (13) implies that the
neutralino
is mostly a Bino in the scaling limit.  However, a note of caution is needed in
that one should not take the $|M_Z/\mu|\rightarrow\infty$ limit.  This limit is
dangerous since the
gaugino-higgsino interference terms which are proportional to
$(n_1,n_2).(n_3,n_4)$
vanish in this limit.  In realistic analyses, as in the computation of the
coherent part of neutralino-nucleus scattering, such terms make significant
contributions and cannot be set to zero.

\section{Neutralino Detection via Neutralino - Nucleus Scattering}
Various possibilities for the detection of neutralino dark matter have been
discussed in the literature.  For example, annihilation of neutralinos in the
galactic halos can produce an observable signal, i.e.,
$\chi+\chi\rightarrow A+X$,
where A  can be an energetic gamma ray, positron or an anti-proton$^{4}$.
  However,
the backgrounds in these processes are rather significant so this process does
not seem very encouraging for the detection of the neutralino.  A second
possibility is that the halo neutralinos are captured in the center of
earth
and sun, annihilate and produce upward moving neutrinos (and muons) in
detectors
on earth$^{14,15,16}$.  The background in these processes are significantly
reduced due to
the angular windows around earth and sun.  Current estimates, however, show
that
one needs around $10^3-10^4 m^2$ neutrino telescopes to see any significant
effect.
The telescopes currently being planned aim to approach $O(10^3 m^2)$ area.
So once
again this mode of detection also does not appear very optimistic for the
neutralino dark matter.  The most optimistic mode for detection of neutralinos
appears to be scattering of neutralinos off nuclei in cryogenic
detectors$^{1,4,19,13,14,17,18}$.
Several detectors using this mode are in various stages of development.  We
shall focus on this mode of detection for the rest of the talk.  More recently
there has also been a discussion of detection of neutralinos via atomic
excitations$^{20}$, but at the moment this possibility requires more
investigation.
The prime detector in neutralino-nucleus scattering is the quark and the
effective interaction that governs the neutralino-quark scattering consists
of a spin-dependent(incoherent) part and a spin-independent (coherent) part,
and
is given by$^{4}$

\begin{equation}
{\cal L}_{eff} = \bar{\tilde Z_1}\gamma_\mu\gamma_5\tilde Z_1 \bar q
\gamma^\mu
(A_q P_L + B_q P_R)q + \bar{\tilde Z_1}\tilde Z_1
m_q
\bar q C_q q
\end{equation}
\noindent
Here $A_q, B_q$  are the spin-dependent amplitudes which arise from the
s-channel
Z-exchange and the t-channel squark exchange, and $C_q$ is the spin-independent
amplitude arising from the s-channel Higgs exchange and t-channel squark-
exchange.  Realistically, of course the quarks are bound inside nuclei so a
reasonable amount of nuclear physics enters in the anlaysis.  The total event
rate is given by$^{4}$

\begin{equation}
R = [R_{coh}+R_{inc}]~\left[{\rho_{\tilde Z_1}\over 0.3~{\rm
GeV~cm}^{-3}}\right]~
\left[{\langle v_{\tilde Z_1}\rangle \over 320~{\rm km/s}}\right]~{{\rm
events}\over
{\rm kg~da}}
\end{equation}

\begin{equation}
R_{coh} = {16m_{\tilde Z_1} M^3_N M^4_Z\over [M_N + m_{\tilde Z_1}]^2}~
\vert A_{coh}\vert^2
\end{equation}

\begin{equation}
R_{inc} = {16m_{\tilde Z_1} M_N \over [M_N + m_{\tilde Z_1}]^2}~
\lambda^2 J(J+1)~\vert A_{inc}\vert^2~,
\end{equation}

\noindent
In the above $\rho_{\tilde Z_1}$  is the local density of dark matter
$v_{{\tilde Z}_1}$ is the relative
velocity, $M_N$ is the nucleus mass, J is the nucleus spin and $\lambda$
 is defined by
%
$ \lambda <N|\vec J|N>=<N|\Sigma \vec S_i|N>$
 where $S_i$  is the nucleon spin.  From eqs. (18)
and (19) one finds that for large $M_N$ one has
$R_{coh}\sim O(M_N)$,
$R_{inc}\sim O(\lambda^2 J(J+1)/M_N)$.
Thus in principle one has two qualitatively different types of detectors, i.e.,
those with large $M_N$ and those with large values of
$\lambda^2J(J+1)$.  In practice there is
seldom a case where $R_{coh}$ is negligible and realistic analyses even for
light
target material (e.g.,$^3$He,CaF$_2$) require inclusion of both $R_{inc}$
and $R_{coh}$.
For heavy targets (e.g., Ge,NaI,Pb) $R_{inc}$ is typically small, i.e. only a
few percent of the total R.

We note that there are uncertainties in the evaluation of both $R_{inc}$ and
$R_{coh}$.
Uncertainties in $R_{inc}$ arise due to experimental uncertainties in the
determination of $\Delta q$ on which $R_{inc}$ depends, where $\Delta q$
are the quark
polarizabilities defined by $<p(n)|\bar q\gamma_{\mu}\gamma_5q|p(n)>=
S^{p(n)}_{\mu}\Delta q$, where $S^{p(n)}_{\mu}=(0,\vec S^{p(n)})$ is the
nucleon spin.  We shall discuss later the sensitivity of the results to the
determination of $\Delta q$.  There is also an uncertainty in the determination
 of $R_{coh}$.
This arises due to the uncertainty in the determination of s-quark matrix
elements$^{21}$
 $<n|m_s\bar ss|n>=f_s M_n$ that enter in the computation of $C_s$.
Currently $f_s$ has a significant uncertainty, about $50\%$, which can lead to
an
uncertainty of $0(30-50)\%$ in the determination of $R_{coh}$.

Next we discuss briefly the relative contribution of the heavy neutral Higgs to
the coherent part of the scattering.  Naively one might expect the heavy Higgs
contribution to be neglible since it would be suppressed by the heavy Higgs
$(mass)^2$ while the light Higgs contribution is suppressed only by the light
Higgs $(mass)^2$.  However, this assessment is correct for the up quark but not
for
the down as can be seen by the expression for $C_q$ below:

\begin{equation}
C^{Higgs}_q = {g^2_2\over 4M_W}~ \left[\left\{
\matrix{{\cos\alpha\over\sin\beta} & {F_ h\over m^2_h} \cr
-{\sin\alpha\over\cos\beta} & {F_h\over m^2_h}}\right\} +
\left\{
\matrix{{\sin\alpha\over\sin\beta} & {F_H\over m^2_H}\cr
{\cos\alpha\over\cos\beta} & {F_H\over m^2_H}} \right\}\right]^{\rm
u-quark}_{\rm
d-quark}
\end{equation}
\noindent
where
\begin{eqnarray}
F_h=(n_1-n_2 tan\theta_W)(sin\alpha n_3+cos\alpha n_4)\\
F_H=(n_1-n_2 tan\theta_W)(-cos\alpha n_3+sin\alpha n_4)
\end{eqnarray}
\noindent
Here $\alpha$ is the Higgs
mixing angle.
Now the Higgs mixing angle $\alpha$ is typically small so from eq. (20) one
finds
that there is a suppression of the d-quark contribution in the light Higgs
sector which often can be more than the $(mass)^2$ suppression in
the heavy Higgs
sector.  Thus the heavy Higgs contribution cannot be neglected as it can make
a substantial contribution to $C_q$.

\section{$b\rightarrow s\gamma$ Branching Ratio Constraint on Event Rates}
Recently the CLEO Collaboration obtained the first experimental determination
of the photonic penguin  process $b\rightarrow s\gamma$.
 For the inclusive $b\rightarrow s\gamma$
decay they find the result$^{22}$

\begin{equation}
BR (b\rightarrow s\gamma ) = (2.32 \pm 0.51 \pm 0.29 \pm 0.32)\times
10^{-4}~
\end{equation}

\noindent
Now in the SM  $b\rightarrow s\gamma$
 decay receives contributions from the
%
%
%
 W-exchange.  For the supersymmetric case
there are additional contributions arising from the exchange of
$H^{+},\tilde W,\tilde Z,and~ \tilde g$.
There are many uncertainties in the theoretical evaluation of
$b\rightarrow s\gamma$. These uncertainties arise from uncertainties in the
determination of $\alpha_s$, quark masses KM matrix elements, and
uncertainties arising from the next to leading order (NLO) QCD corrections
which can be as large as $0(30)\%$.  There is a significant debate in the
literature currently regarding what exactly the $BR(b\rightarrow s\gamma)$
 value is in the SM$^{23}$.
An accurate answer to this question can only result after the next-to-leading
order(NLO) QCD corrections
have been computed reliably.  Similar uncertainties are present in the
evaluation of $b\rightarrow s\gamma$ in SUSY theory.  The discussion of
$b\rightarrow s\gamma$ constraint on neutralino relic density $^{24,25}$
can be facilitated by use of the parameter
$r_{SUSY}$$^{18}$ which we
define by the ratio
%
$r_{SUSY} = BR(b\rightarrow s\gamma)_{SUSY}/BR(b\rightarrow
s\gamma)_{SM}~$.
Many of the uncertainties discussed above  cancel out in
the ratio $r_{SUSY}$.  However, we must keep in mind that the NLO corrections
 would
in general be different for the SUSY case than for the SM case.  In this
analysis, however, we limit ourselves to the leading order evaluation.
Analogous to $r_{SUSY}$  we can also define$^{18}$
%
$r_{expt} = BR(b\rightarrow s\gamma)_{expt}/BR(b\rightarrow
s\gamma)_{SM}~.$
The CLEO Collaboration use an SM value of $(2.75\pm 0.8)\times 10^{-4}$.
Using this value and the result of eq. (23) one finds that $r_{exp}$ lies
in the range
$r_{exp}=0.46-2.2.$
This range turns out to be a rather strong constraint on SUSY theory if we
assume
that $r_{SUSY}=r_{exp}$$^{18}$.  This is so because in the SUSY case one
normally gets
a much larger range for $r_{SUSY}$ , i.e., a range of $\approx (0,10)$.

\section{Analysis and Results}
We first discuss the event rates without inclusion of the $b\rightarrow
s\gamma$
constraint.  Results on maximum and minimum of event rates are exhibited in
Fig. 1 as a function of $m_{gluino}$ for the target materials:
Pb, Ge and
CaF$_2$.
We note that the maximum curves of the event rates all exhibit a dip when
$m_{gluino}$ is in the mass range $\approx 300-450 GeV$ independent of the
target
material.  This dip arises$^{18}$ due to the Z-pole and the Higgs pole effects
and the
relic density constraint.  Effectively, rapid annihilation near the Z and Higgs
poles leads to values of $\Omega h^2$ which fall below the CHDM limit and have
to be
eliminated.  The eliminated part of the parameter space contains the light SUSY
spectrum and the large event rates.  Thus one sees a dip in the event rates in
the vicinity of $m_{\tilde Z_1}\sim M_Z/2$ and
 $m_{\tilde Z_1}\sim m_h/2$.
The total event rate R is generally dominated by $R_{coh}$ for all except the
lightest target materials.  Now $R_{coh}$ depends on the gaugino-higgsino
interference term which is proportional to $(n_1,n_2)\times
(n_3,n_4)$.
Using the behavior of $n_i$ with large $\mu$ and the dependence of $R_{coh}$
on the
gaugino-higgsino interference, we can understand the behavior of R for large
$m_{gluino}$.  Typically $\mu$ is an increasing function of $m_{gluino}$,
and  the $n_i$(i=1,3,4) are decreasing
functions as $\mu$ increases.  Thus $R_{coh}$ falls monotonically$^{18}$ with
$m_{gluino}$ beyond the dip
as can be seen in Figs. 1.  (A similar analysis holds as a function of $m_0$).
 Now
it is easily seen that for $m_{gluino}\geq 650 GeV$ for $\mu<0$ and
$m_{gluino}\geq700 GeV$ for $\mu>0$  the event
rates even for heavy targets (e.g. Pb) fall below the level of 0.01 event/kg.d.
This is the level of sensitivity that detectors in the next 5-10 yrs hope to
achieve.
\begin{figure}
\vspace*{13pt}
\leftline{\hfill\vbox{\hrule width 5cm height0.001pt}\hfill}
\vspace*{1.4truein}             
\leftline{\hfill\vbox{\hrule width 5cm height0.001pt}\hfill}
\fcaption{Plot of maximum and minimum event rates
as functions of $m_{gluino}$ for
CaF$_2$(dash-dot),
Ge(dash) and Pb(solid) when $tan\beta\leq20$  and other parameters span the
parameter space..
Fig. 1a is for $\mu< 0$ and Fig. 1b for $\mu >0$.}
\label{fig:radk}
\end{figure}

Next we discuss the effect of the $b\rightarrow s\gamma$ constraint on
dark matter$^{24,18,13}$.  The
quantitive effects of $b\rightarrow s\gamma$ depend on the value of
$r_{SUSY}$.  Generally, the
$b\rightarrow s\gamma$ constraints is more severe for $\mu >0$ than for
$\mu <0$.  Similarly, one finds
a significant effect on the event rates for $\mu >0$, while the effect for
$\mu <0$ is
relatively smaller.  Results for the case $r_{SUSY}\leq1.33$ is shown in Fig. 2
for the
cases $\mu >0$ (Fig. 2a) and $\mu <0$(Fig. 2b).  As indicated above the allowed
region
of the parameter space shrinks significantly for $\mu >0$ (Fig. 2a) and the
maximum
allowed event rates also fall. The corresponding effect on $\mu\leq0$ (Fig. 2b)
is
signficantly smaller.

\begin{figure}
\vspace*{13pt}
\leftline{\hfill\vbox{\hrule width 5cm height0.001pt}\hfill}
\vspace*{1.4truein}             
\leftline{\hfill\vbox{\hrule width 5cm height0.001pt}\hfill}
\fcaption{Same as Fig 1 except that the $b\rightarrow s\gamma$
 constraint of $r_{SUSY}\leq1.33$
is imposed.}
\label{fig:radk}
\end{figure}

Finally we discuss the effect of the variations in quark polarizabilities
$\Delta q$ on
the event rates.  The part sensitive to $\Delta q$ is $R_{inc}$.
The previous
determinations of $\Delta q$ using the EMC data gave$^{26}$,
$(\Delta u,\Delta d,\Delta s)=(0.77\pm0.08,-0.49\pm0.08,-0.15\pm0.08)$.
Recently there has been a reanalysis of $\Delta q$ using new data from
SMC$^{27}$ which gives$^{28}$
%
$(\Delta u,\Delta d,\Delta s)=(0.83\pm0.03,-0.43\pm0.03,-0.10\pm0.03)$.
These determinations are consistent with each other within
$1\sigma$, but the variations of $\Delta q$, specifically the variation of
$\Delta s$,
can generate signficant changes in $R_{inc}$.  However, $R_{inc}$ is
generally a small
component of the total R$^{18,13}$.  We have analysed several target materials,
He,CaF$_2$,Ge,Pb etc., and find that except for the lightest target
materials, i.e. He and CaF$_2$, $R_{inc}$  is only a few precent of the the
total.
Thus the event rates for targets such as Ge are not significantly
affected by the  new determination of $\Delta q^{18,13}$.

\section{Concluding Remarks}
We have given here an analysis of neutralino dark matter within
minimal supergravity unification.
Remarkably the model predicts  that the lightest neutralino is also the LSP
over
most of the parameter space of the model, and thus the model gives a
candidate for cold dark matter.  We have analysed the relic density of
neutralinos, using the accurate method which integrates
over the Z and Higgs poles in thermal averaging of $\sigma v$.  These effects
are
found to be significant for values of gluino masses $\leq 400GeV$.  We have
analysed the event rates in neutralino-nucleus scattering and find that there
is
a significant region of the parameter space where event rates $\geq 0.01$
 event/kg.d
are predicted for targets such as Ge and Pb.  This region of the parameter
space
will be accessible to current and future technologies over the next 5-10 years.
However, more sensitive detectors, two to three orders of magnitude
 more sensitive,are needed  to probe most of the parameter space of
the minimal supergravity model.

\section{Acknowledgements}
This research was supported in part by NSF grant numbers PHY-19306906,
PHY-9411543, and PHY94-07194.

\end{document}